\documentclass[letterpaper]{article}

\usepackage{natbib,alifeconf}  %% The order is important

\usepackage{diagbox}
\usepackage[hidelinks]{hyperref}
\usepackage{multirow}
\usepackage{tabu}
\usepackage{lipsum}
\usepackage{amssymb}
\usepackage{amsmath}
\usepackage{graphicx}
\usepackage{dblfloatfix} 
\title{Emergent Escape-based Flocking Behavior\\ using Multi-Agent Reinforcement Learning}
\author{Carsten Hahn$^{1}$, Thomy Phan$^{1}$, Thomas Gabor$^{1}$, Lenz Belzner$^{2}$ and Claudia Linnhoff-Popien$^{1}$ \\
\mbox{}\\
$^1$Mobile and Distributed Systems Group, LMU Munich, Munich, Germany \\
$^2$MaibornWolff, Munich, Germany \\
carsten.hahn@ifi.lmu.de} % email of corresponding author

\begin{document}

\maketitle

\begin{abstract}
% Abstract length should not exceed 250 words
In nature, flocking or swarm behavior is observed in many species as it has beneficial properties like reducing the probability of being caught by a predator.
In this paper, we propose SELFish (Swarm Emergent Learning Fish), an approach with multiple autonomous agents which can freely move in a continuous space with the objective to avoid being caught by a present predator. 
The predator has the property that it might get distracted by multiple possible preys in its vicinity.
We show that this property in interaction with self-interested agents which are trained with reinforcement learning to solely survive as long as possible leads to flocking behavior similar to Boids, a common simulation for flocking behavior.
Furthermore we present interesting insights in the swarming behavior and in the process of agents being caught in our modeled environment.

\end{abstract}

\section{Introduction}
Flocking or swarm behavior is observed in many species in nature. 
A prominent example is fish schooling, where multiple fishes do not only stay close to each other for social reasons but coordinate their actions collectively. 
That means that an individual fish aligns its direction in regard to fishes that are close to it, while maintaining a certain cohesion of the group and still avoiding collisions with other individuals.

However, flocking behavior does not only exist as an end in itself.
In nature, a schooling fish benefits from schooling in multiple ways:
The swarm increases one's hydrodynamic efficiency or mating chances.
Also, flocking enhances foraging success as collaborative observation is superior to a single individual's.
The same is true for predator detection.
Even further, the probability of being caught decreases for an individual with regard to certain predator behaviors.

\cite{Boids} showed that algorithmically implementing the three rules of alignment, cohesion and separation leads to flocking behavior while an individual only needs local knowledge about its surrounding neighbors (called Boids).
In order to overcome these static flocking rules \cite{RLSwarm} used reinforcement learning to train an individual to justify the rules stated above in order to form a swarm.
This was done by shaping the reward signal according to distances between the individuals and limiting their actions to be attracted to another fish, be repulsed from another fish and move parallel in the same or opposite direction of another fish, respectively.

With SELFish we investigate the case that an individual tries to optimize its behavior with respect to the objective of surviving as long as possible in the presence of a predator (which might get distracted by multiple preys).
We show that this simple objective leads to emergent flocking behavior (similar to Boids) in a multi-agent reinforcement learning setting, without the need to explicitly enforce it.

\section{Reinforcement Learning}

Reinforcement Learning denotes a machine learning paradigm in which an agent interacts with its environment and receives a certain reward for its action accompanied with an observation of the new state of the environment. 
Such scenarios are usually modeled as Markov Decision Processes (MDPs), where $ \mathcal{S} $ denotes the set of states of the environment, $ \mathcal{A} $ denotes the set of actions an agent can take and $ r(s_t, a_t) $ is the intermediate reward received after action $ a_t $ was taken in state $ s_t $ at time step $ t $.
Also, the process moves to a new state $ s_{t+1} $ influenced by the action $ a_t $, with the Markov property being that the new probability of transitioning into state $ s_{t+1} $  only depends on state $ s_t $ and the chosen action $ a_t $: $ \mathcal{P}(s_{t+1}|s_t, a_t) $.
The goal is to find a policy $ \pi: \mathcal{S} \rightarrow \mathcal{A}$ which maximizes the accumulated reward $ R_t = \sum_{i=t}^{T} \gamma^{i-t} r(s_i, a_i)$ from time step $ t $ to the simulation horizon $ T $ with a discounting factor $ \gamma \in [0,1]$.

In SELFish the state is partially observable, which means that instead of using the full state description $s_t$ to determine the action $a_{t} = \pi(s_t)$, the agent only uses an observation $o_t \in \mathcal{O} $ (where $\mathcal{O}$ is the space of all possible observations) as input to a policy function $\pi : \mathcal{O} \rightarrow \mathcal{A}$ to compute the action $a_{t} = \pi(o_t)$.
Furthermore the observation may be different for every agent.
However, we focus on a deterministic domain, so $ \mathcal{P}(s_{t+1}|s_t, a_t) \in \{0, 1\} $.

\subsection{Deep Learning}
In Reinforcement Learning the policy or intermediate functions, which help to derive it, are usually expressed as deep artificial neural networks.
Neural networks can viewed as a directed graph of nodes, called neurons, which are interconnected by weighted edges. 
A neuron receives inputs over its ingoing edges, usually computes the weighted sum of the inputs, applies a non-linear function to this weighted sum and forwards its output to subsequent neurons via its outgoing edges.
The neurons are usually arranged in layers, where layers between the input layer and the output layer of the network are referred to as hidden layers.
Networks with multiple hidden layers are called deep neural networks.

Artificial neural networks serve as biologically inspired function approximators which can be trained by example to approximate a function $ f $ mapping an input vector $ x \in {\rm I\!R}^n $ to an output vector $ y \in {\rm I\!R}^m $ depending on the weights of the edges $ \theta $.
The goal in training a neural network is to minimize the error between the networks' output $ y' = f(x; \theta) $ and the known desired (example) output $ y $ by adjusting the weights $ \theta $ accordingly. 
This can be done with the Backpropagation method combined with a gradient descent strategy.

\subsection{Deep Q-Learning (DQN)}
Q-Learning is a value-based approach named after the action-value function $ Q^\pi: \mathcal{S} \times \mathcal{A} \rightarrow {\rm I\!R} $, which describes the expected accumulated reward $ Q^\pi(s_t, a_t)$ after taking action $ a_t $ in state $ s_t $ and following the policy $ \pi $ in all subsequent states.
The goal is to find an optimal action-value function $ Q^* $, which yields the highest accumulated reward.
$ Q^* $ can be approximated through Bellman's principle based on the intuition that for an optimal policy, independently of the initial state and initial decision, all remaining decisions must constitute an optimal policy with regard to the state resulting from the first decision (\cite{bellman1957dynamic}).
Starting from an initial guess for $ Q $, it can be iteratively updated via
$$Q(s_t, a_t) \leftarrow Q(s_t, a_t) + \alpha [r_t + \gamma \max_a Q(s_{t+1}, a) - Q(s_t, a_t)]$$
where the learning rate $ \alpha \in (0, 1)$ is a parameter to be specified.
The learned action-value function $ Q $ converges to $ Q^* $, from which an optimal policy can be derived via $ \pi^*(s_t) = \arg\max_a Q(s_t, a) $.

In Deep Q-Learning (DQN) (\cite{DQN}) an artificial neural network is used to represent the action-value function $ Q $. 
Also, to minimize correlations between samples and to alleviate non-stationary distributions an experience replay mechanism is used (\cite{DQN}) which randomly samples previous state action transitions to train the neural network.

\subsection{Deep Deterministic Policy Gradient (DDPG)}
To overcome the limitation of Q-Learning,  which cannot directly be applied to continuous action spaces, efforts were made to learn the policy $ \mu(s|\theta^\mu) $ directly with a parameterized objective function $ J(\theta) $ (\cite{DPG,DDPG}).
In addition it was proposed to split the learning process in two components to reduce the gradient variance, called actor-critic approach.
The critic learns the action-value function $ Q(s, a) $ using the Bellman equation as in Q-learning.
The actor then updates the policy parameters $ \theta^\mu $ in the direction suggested by the critic:
\begin{equation*} 
\nabla_{\theta^\mu}J = \mathbb{E}_{s_t}[\nabla_a Q(s,a|\theta^Q)|_{s=s_t, a=\mu(s_t)} \nabla_{\theta^\mu \mu(s|\theta^\mu)|_{s=s_t}}]
\end{equation*}

\subsection{Multi-Agent Case}
Many approaches have been suggested for the case that there are multiple agents present which are either self-interested or have to work together to achieve a cooperative goal.
A straightforward idea in the case that there are multiple agents that act in their self-interest, which means that they only maximize their own accumulated reward, is deploying a standard reinforcement learning algorithm (as in the single-agent case) in each individual agent in the multi-agent setting and let all agents learn simultaneously.
This straightforward approach bears the problem of non-stationarity in the state transitions.
As one agent tries to adapt its actions in certain states, other agents, which are considered as part of the environment for the first agent, do so as well.
This makes it difficult to learn a policy depending on the observed state, which no longer satisfies the Markov property.

\cite{MultiAgentDQN} approaches a pursuit-evasion game with reinforcement learning.
There are multiple pursuers and multiple evaders.
Only one agent of each kind is trained through Q-Learning at a time while the policies of the other agents are fixed.
After a number of iterations the policy of the learning agent is distributed to all other agents of the same type.
Through this process the policy of one set of agents is improved incrementally over time.

This mitigates the problem of non-stationarity.
Furthermore it seems reasonable to copy the policy of one agent throughout multiple homogenous agents as all are alike and pursue the same self-interested goal.
This observation is also relevant for flocking or swarm behavior of multiple agents as we will demonstrate below.

\section{Swarm Behavior}
In 1987, Craig Reynolds (\cite{Boids}) described three basic rules through which flocking behavior can be modeled.
For these rules an individual only needs local knowledge about its neighbors within a certain distance.
These rules are:
\begin{itemize}
\item \textbf{Alignment:} Steer towards the average heading direction of local flockmates
\item \textbf{Cohesion:} Steer towards the average position (center of mass) of local flockmates
\item \textbf{Separation:} Steer to avoid crowding local flockmates
\end{itemize}
If each individual (called \textit{Boids} by Reynolds as he thought of bird-like creatures) follows these rules, a swarm formation emerges.
In an implementation, they can be expressed as physical forces which act upon an individual. 
Supplementary forces can be introduced, which repel an individual from an enemy or from obstacles, for example.

To overcome these static rules definitions, \cite{RLSwarm} used Reinforcement Learning, particularly Q-Learning, to train agents to follow these rules.
In their model the agents iteratively learn while at every time step an agent $ i $ only considers one other agent $ j $.
Agent $ i $ receives the euclidean distance to $ j $ as observation and can choose among four actions to execute.
These actions are to move towards agent $ j $, away from agent $ j $ or parallel to agent $ j $  either in the same or opposite direction.
The reward agent $ i $ receives for an action depends on the previously mentioned distance to agent $ j $ and is shaped in a way that it intuitively represents the cohesion and separation rule.
In this regard agent $ i $ receives a positive reward if it steers so to keep its distance to $ j $ within predefined boundaries.

While the previously mentioned approaches lead to flocking behavior, they neglect the beneficial properties flocking behavior might have for the individuals.
One of those benefits could be the increased likehood to survive in the presence of predators, as they might get distracted by the sheer amount of possible targets.
The question arises whether flocking behavior occurs in a scenario with such properties where agents solely try to maximize their survival time.
In contrast to \cite{RLSwarm}, we pursue a scenario in which agents are trained with reinforcement learning solely on the objective to survive, without explicitly enforcing swarm behavior.
Additionally, we demonstrate that SELFish also works for a continuous action space of the agents.

\section{Emergent Swarm Behavior}
In order to investigate whether the objective to survive in the presence of a predator would lead to flocking behavior in a multi-agent setting, we created a model that facilitates such a behavior. 
In the following the properties of the environment will be explained.
This is followed by a description of the action and observation space as well as the reward structure which was used to train the agents.

\subsection{Environment}
\begin{figure}[t]
\begin{center}
\includegraphics[width=.25\textwidth]{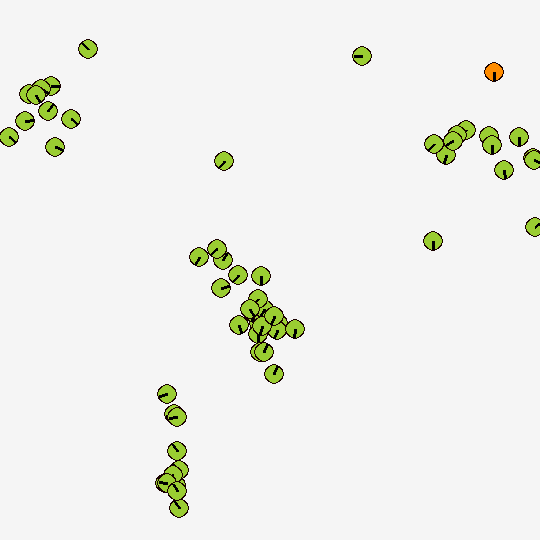}
\caption{Example of the space with $ 60 $ agents (green) and one predator (orange).}
\label{fig:space}
\end{center}
\end{figure}
The agents, which are the prey in this scenario, can freely move in a continuous two-dimensional space, visualized as a square with predefined edge lengths (see Figure \ref{fig:space}). 
An agent itself is represented as a circle with a surface substantially smaller than the space it is moving in.
There are neither obstacles nor walls in the environment. 
Furthermore agents do not collide with each other.
To ease free roaming of the agents, the space has the special characteristic that it wraps around at the edges forming a torus. % (see Figure \ref{fig:torus}).
That means that if an agents leaves the square visualization to the right, it will immediately enter it again from the left (same for the other direction or around top and bottom). 

%\begin{figure}[t]
%\begin{center}
%\includegraphics[width=0.3\linewidth]{fig/torus.pdf}
%\caption{Sketch of environment wrapping around as a torus.}
%\label{fig:torus}
%\end{center}
%\end{figure}

Together with the agents there also exists a predator in the environment.
The predator is also represented as a circle.
The goal of the predator is to catch the agents by moving to their position.
As soon as the predator collides with an agent, the execution of the concerning agent will end and a new agent is  spawned immediately at a random position to keep the number of agents in the system constant.
If there are multiple agents within a certain distance around the predator, it will choose one for a target at random (otherwise it will move to the closest agent's direction). 
This means that the predator can be distracted by multiple agents in its proximity. 
Thus it might be beneficial for an agent to move towards other agents as the predator might get distracted, which is essential for flocking behavior.
However, to prevent the predator from constantly changing targets it will follow a chosen target for a certain time before a new target will be chosen.
By default, the agents and the predator move at the same speed.
This would allow an agent to turn in the opposite direction of the predator and move away without the predator having a chance to catch up.
That is why the predator will accelerate occasionally for a short amount of time, which simulates a leap forward to catch the prey it is following.
The policy of the predator is static and does not change over time.

\subsection{Objective of an Agent}
The goal of the agents is not to collide with the predator.
For this they receive a reward of $ \textbf{+1} $ for each step/frame they live and $ \textbf{-1000} $ for the collision with the predator which ends their life.
With this reward structure the objective of the agents can be viewed as ``surviving as long as possible''.
As there are no obstacles in the environment and the agents do not collide with each other, there are no other rewards/penalties.

\subsection{Action Space}
The action space of the agents only comprises of the angle they want to turn each time step.
The movement speed of the agents is constant and cannot be altered by them for now.

The action $ a $, which represents the turning angle that can be chosen from discrete steps or out of a continuous interval by the agent, depends on the reinforcement learning strategy which is used later on.
In the case that DQN is used, the actions an agent can chose from comprise five discrete degree values $\{-90^{\circ}, -45^{\circ}, 0^{\circ}, +45^{\circ}, +90^{\circ}\}$. %$ \{ a \in \mathbb{Z} \;| -\!90^{\circ} \leq a \leq 90^{\circ} \wedge \, a \mod 45^{\circ} = 0  \} $. 
The agent can choose any real-valued degree as turning angle in the case of DDPG.

As a side node, the predator can only take limited real-valued turns $\{ x \in {\rm I\!R} \;| -45^{\circ} \leq x \leq 45 ^{\circ} \}$  at every step with the goal to give the agents a higher maneuverability than the predator.

\subsection{Observation Space}
In order to facilitate the scalability to many autonomous agents, one agent cannot observe the full state of the environment; instead its observation is limited to itself, the predator and the $ n $ nearest neighboring agents.
This approach can be explained biologically, where, for example, a fish in a swarm cannot observe the whole swarm but only its local neighbors.
But it is also in line with related work, for example Boids, where also only local neighborhoods between agents are regarded.
Furthermore it eases computation and has the nice property that the observation vector, which is forwarded through the reinforcement learning algorithm in order to obtain an action, has a constant length (cf. the following section).

For every observable entity $ e $, the agent receives a 3-tuple which contains the euclidean distance between the entity and the agent, the angle the agent would have to turn to face towards the observed entity and the absolute orientation of the entity in the environment: $ (\textit{dist}_e, \textit{direction}_e, \textit{orientation}_e) $.
As the environment is a torus, the distances are also calculated around the edges of the visualized square, with the shorter distance being taken (with the $ \textit{direction}_e $ corresponding to this). 
The absolute orientation of an entity is measured in degrees $ [0^\circ,360^\circ) $, where facing east corresponds to $ 0^\circ $, measuring the angle counter-clockwise.
The angle an agent would have to turn to face towards another entity is measured in degrees in the range of $ (-180^\circ,180^\circ] $.

Accordingly, an agent receives the following observation for the predator, itself and the $ n $ nearest neighboring agents, in which the $ n $ neighbors are ordered by their distance.
$$\left[
\begin{array}{ccc}
\textit{dist}_\textit{predator} & \textit{direction}_\textit{predator} & \textit{orientation}_\textit{predator}\\ 
0 & 0 & \textit{orientation}_\textit{self}\\
\textit{dist}_{\textit{neighbor}_1} & \textit{direction}_{\textit{neighbor}_1} & \textit{orientation}_{\textit{neighbor}_1}\\
\textit{dist}_{\textit{neighbor}_2} & \textit{direction}_{\textit{neighbor}_2} & \textit{orientation}_{\textit{neighbor}_2}\\
 & \vdots & \\
\textit{dist}_{\textit{neighbor}_n} & \textit{direction}_{\textit{neighbor}_n} & \textit{orientation}_{\textit{neighbor}_n}
\end{array}
\right]
$$

\subsection{Training}\label{sec:training}
As mentioned before, a valid way for training multiple homogeneous agents through reinforcement learning is to train only one instance and then to copy the learned policy to all instances of the homogeneous group (\cite{MultiAgentDQN}).
This also resembles nature, where for example multiple schooling fish follow the same behavioral policy.

For this purpose, the DQN and DDPG implementations of Keras-RL (cf. \cite{keras-rl}) were used.
Keras-RL is originally developed for OpenAI Gym Environments (\cite{keras-rl}), in which only single agents interact with these environments through a \textit{step(action)}-method, which is given an action and returns an observation, a reward and a done flag, indicating whether the current episode is finished.
This interface was also used in the proposed swarm environment to train a single agent to avoid the present predator with the previously mentioned rewards, action and observation spaces.
During the training of one agent, the other agents are present as well, onto which the policy (i.e. the neural network) of the learning agent is copied after each episode.
An episode ends if the learning agent is caught by the predator or $ 10,000 $ steps (frames) were executed.

During training, the edge lengths of the space were $ 40 \times 40 $ pixels, although it wraps around at the edges.
Please note that the agents and the predator could be positioned at any real value in the interval $ [0, 40] $.
However, the values in the 3-tuples of the observation were normalized to $ [0,1] $ anyway.
The agents and the predator were represented by circles of radius $ 1 $, with an agent being caught if the distance of its position and the position of the predator is below $ 2 $.
Also, during training only $ 10 $ agents were present.

\begin{table}[t]
\begin{tabular}{lrr}
%\multicolumn{1}{c}{}                                   & \multicolumn{2}{c}{Value}                                                                                                                         \\
\multicolumn{1}{c|}{Hyperparameter}                    & \multicolumn{1}{c|}{DQN}                 & \multicolumn{1}{c}{DDPG}                                                                               \\ \hline
\multicolumn{1}{l|}{Training Steps}                    & \multicolumn{1}{r|}{500,000}             & 500,000                                                                                                \\
\multicolumn{1}{l|}{Hidden Layer}                      & \multicolumn{1}{r|}{10}                  & 5                                                                                                      \\
\multicolumn{1}{l|}{\multirow{2}{*}{Neurons in Layers}} & \multicolumn{1}{r|}{\multirow{2}{*}{16}} & Actor: 16                                                                                              \\
\multicolumn{1}{l|}{}                                  & \multicolumn{1}{r|}{}                    & Critic: 32                                                                                             \\
\multicolumn{1}{l|}{Hidden Layer Activation}           & \multicolumn{1}{r|}{relu}                & relu                                                                                                   \\
\multicolumn{1}{l|}{Last Layer Activation}             & \multicolumn{1}{r|}{linear}              & linear                                                                                                 \\
\multicolumn{1}{l|}{$\gamma$}                          & \multicolumn{1}{r|}{0.999999}            & 0.999999                                                                                               \\
\multicolumn{1}{l|}{Optimizer}                         & \multicolumn{1}{r|}{Adam}                & Adam                                                                                                   \\
\multicolumn{1}{l|}{Learning Rate}                     & \multicolumn{1}{r|}{0.001}               & 0.001                                                                                                  \\
\multicolumn{1}{l|}{Replay Buffer Size}                & \multicolumn{1}{r|}{50,000}              & 100,000                                                                                                \\
\multicolumn{1}{l|}{Batch Size}                        & \multicolumn{1}{r|}{64}                  & 512                                                                                                    \\
\hline
\multicolumn{1}{l|}{\multirow{2}{*}{Exploration}}      & \multicolumn{1}{r|}{$ \epsilon $-Greedy} & \begin{tabular}[c]{@{}r@{}}Ornstein \\ Uhlenbeck\end{tabular}                                          \\
\multicolumn{1}{l|}{}                                  & \multicolumn{1}{l|}{$ \epsilon = 0.1$}   & \multicolumn{1}{l}{\begin{tabular}[c]{@{}l@{}}$\theta=0.15,$\\ $\mu=0.0,$\\ $\sigma=0.3$\end{tabular}} \\ \hline
\multicolumn{1}{l|}{Observable neighboring agents}     & \multicolumn{1}{r|}{5}                   & 1                                                                                                     
\end{tabular}
\caption{Hyperparameters for Reinforcement Learning}
\label{tab:parameter}
\end{table}

\begin{figure*}[t]
\begin{center}
\includegraphics[width=.9\textwidth]{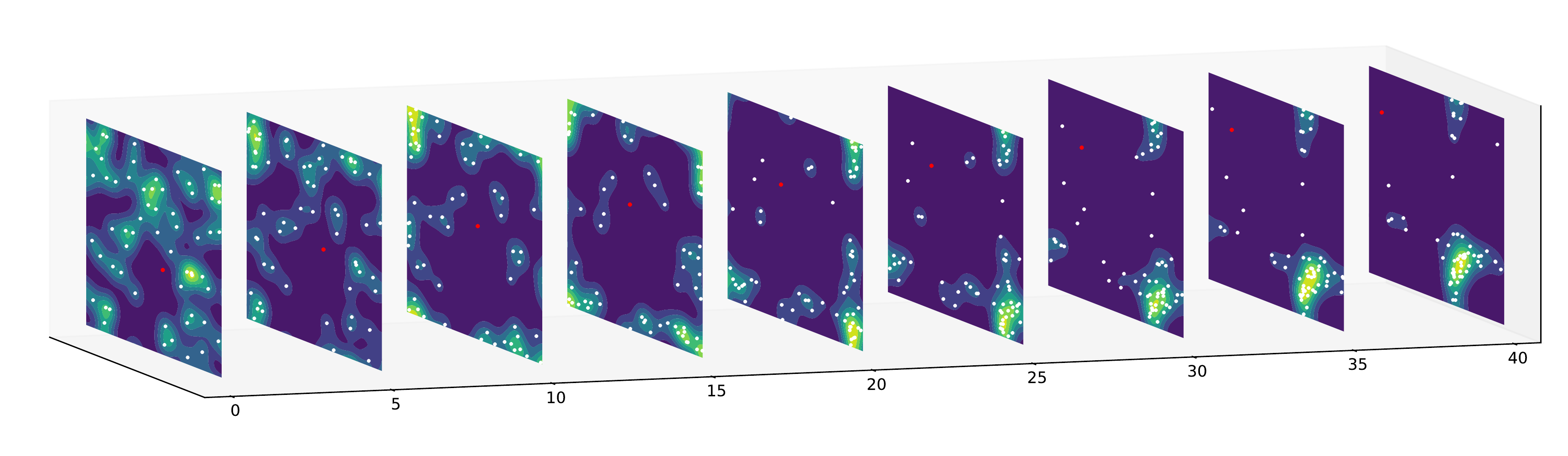}
\caption{Swarm formation in the first 40 frames of an episode of $ \text{SELFish}_\text{DQN} $. Agents (white) and predator (red) were randomly initialized. Kernel Density Estimation \cite{KDE} was used to highlight the dense regions of the multi-agent swarm. Note that the space wraps around the edges.}
\label{fig:DQNSwarmFormation}
\end{center}
\vspace*{-10px}
\end{figure*}

%hyperparameter, parameter sichtbare fische,scalierung, schwarm erkennbar
%As there are multiple parameters for the reinforcement learning algorithms 
In order to find a good configuration for the parameters of the reinforcement learning algorithms,
% like learning rate $ \alpha $, the discount factor $ \gamma $ as well as the number of layers and the number of neurons per layer of the neural networks or the memory size for the experience replay, 
many runs were executed.
The quality of the parameter configuration of the training run was assessed during a test phase based on the cumulative reward the learning agent could acquire, which essentially equals the number of time steps it could survive.
The number of neighboring agents that could be observed was also varied as parameter. % which resulted in $ \textbf{1} $ neighbor for DQN and $ \textbf{4} $ neighbors for DDPG as best configuration.
See Table \ref{tab:parameter} for the best parameters found.

Even for the small number of agents which were present during the training, a swarming behavior could be observed when the learned behavior of one agent was transferred to the others.
Since the observation of an agent is partial and thus limited to the 3-tuple $ (\textit{dist}_{\textit{neighbor}_i}, \textit{direction}_{\textit{neighbor}_i}, \textit{orientation}_{\textit{neighbor}_i}) $ for the $ n $ nearest neighbors, the number of agents as well as the size of the space can be increased without breaking the learned policy. 
With this even better swarming behavior can be observed, which shall be further evaluated in the next section.

\section{Simulations and Results}
%In this section we want to discuss the results of the reinforcement learning and the swarming behavior.
%As Boids was the first algorithm proposed for the algorithmic simulation of swarming behavior, we compare the swarms resulting from multi-agent reinforcement learning to those from Boids.

First we want to give an impression of the swarms that are forming from reinforcement learning.
See Figure \ref{fig:DQNSwarmFormation} for the formation of a swarm in the first 40 frames of a test episode of $ \textit{SELFish}_\textit{DQN}$. 
With a continuous action space, $ \textit{SELFish}_\textit{DDPG} $, exhibits similar behavior although the swarm tends to be more dense.
The swarm presumably forms because one agent learns that the predator might get distracted from it if it stays close to other agents which prolongs its life and thereby its accumulated reward.

\textit{Boids} enforces the alignment, cohesion and separation of neighboring agents.
This can be expressed by vector calculations together with weights which set these three rules in context.
To make the scenario more similar to the reinforcement learning setting, another force which pushes the Boids away from the predator was added (altogether with a weight for this behavior which sets it in context to the other rules).
To find a good configuration for the alignment, cohesion, separation and predator avoidance weight, multiple runs with different parameter setting were executed.
Again, the quality of a setting was evaluated based on the number of time steps a certain boid could survive. %(just like the learning agent in reinforcement learning).

If it is only about the survival of an agent, a simple strategy one could think of is to simply turn in the opposite direction of the predator and to move away from it regardless of the surrounding agents.
This policy will be called \textit{TurnAway} in the following and will be given for comparison\footnote{For a short video showing all implemented policies please refer to \url{https://youtu.be/SY59CYaqWpE}}.

\subsection{Alignment and Cohesion}
\begin{figure}[h]
\begin{center}
\includegraphics[width=.445\textwidth]{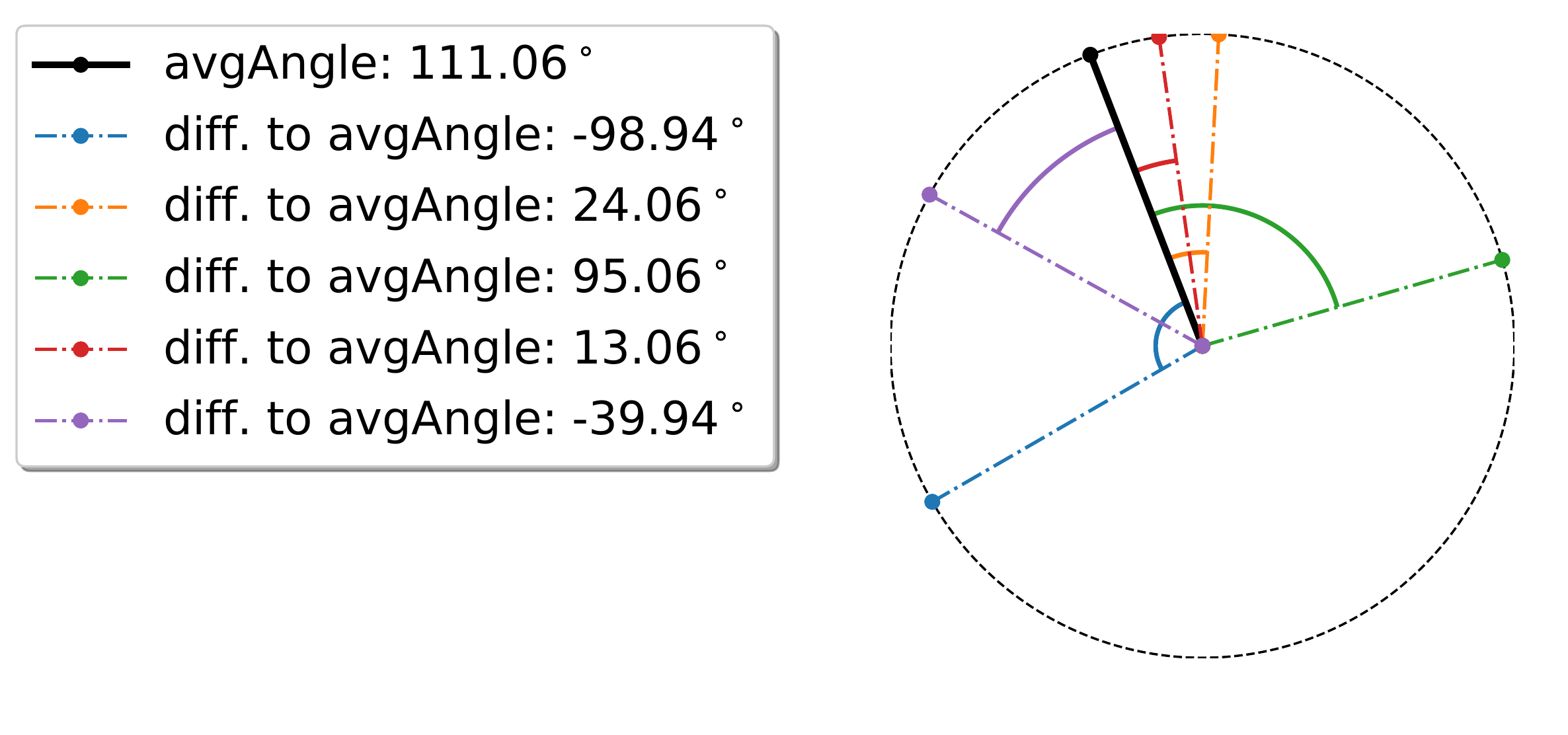}
\caption{Considering the orientation of five agents in space, a mean angle (black) and the deviation from this in $ (-180^\circ, 180^\circ] $ can be computed (\cite{statistics}).}
\label{fig:meanAngle}
\end{center}
\vspace*{-10px}
\end{figure}

As Boids enforces the alignment and the cohesion of the agents, we want to compare the swarms resulting from predator avoidance through reinforcement learning to Boids by these means.
As the orientation of an agent is measured as angle in $ [0^\circ, 360^\circ) $ (facing east corresponds to $ 0^\circ $), the alignment of the agents can be measured as deviation from a mean angle of a group (see Figure \ref{fig:meanAngle}).
The absolute deviation of each agent from this mean angle was summed and averaged over the number of agents.
To measure the cohesion of the swarm, the average distance between the agents was calculated.
For this the distance between all agents $i$ and $j$ %$ \text{agent}_i $ and $ \text{agent}_j $
%\newcommand\bigzero{\makebox(0,0){\text{\Huge0}}}
%$$
%\begin{bmatrix}
%          & 	\textit{dist}_{1,2} & \textit{dist}_{1,3} & \ldots &   \textit{dist}_{1,n} \\
%          &         & \textit{dist}_{2,3} & \ldots &   \textit{dist}_{2,n} \\
%          &         &         & \ddots &    \vdots \\
%          &\bigzero &         &        &   \textit{dist}_{n-1,n}      \\
%          &         &         &        &          
%\end{bmatrix}
%$$
was summed and averaged by the number of pairs of agents.

\begin{figure}[h]
\begin{center}
\includegraphics[width=.25\textwidth]{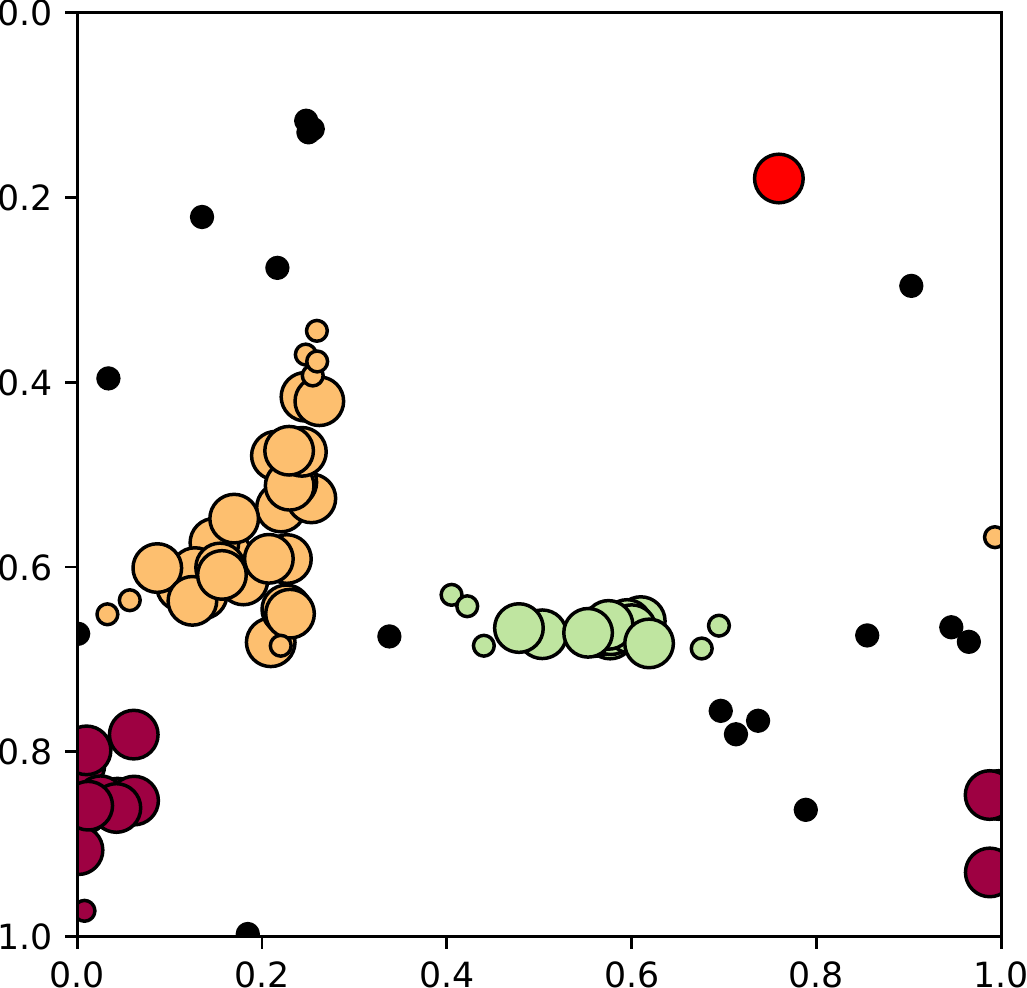}
\caption{Example Clustering for $ \text{SELFish}_\text{DQN} $ with $ 40 $ agents (predator as red dot).}
\label{fig:DQNDBScan.pdf}
\end{center}
\vspace*{-10px}
\end{figure}

Considering that the agents flee from a predator and the space wraps around at the edges, multiple flocks with different orientations, depending on their position in regard to the predator, might form, as it is already evident from the Figures \ref{fig:space} and \ref{fig:DQNSwarmFormation}. 
That is why it did not seem sensible to calculate alignment and cohesion over all agents in the space.
To counter this, the density-based clustering method DBSCAN (\cite{dbscan}) and particularly its scikit-learn implementation (\cite{scikit-learn}) was used beforehand and the average deviation from the mean angle and the average distance between two agents was only calculated for agents in a specific cluster (see Figure \ref{fig:DQNDBScan.pdf} for an example). 
The measurements over all agents are given for comparison. 
\vspace*{-15px}

\begin{figure}[h]
\begin{center}
\includegraphics[width=.46\textwidth]{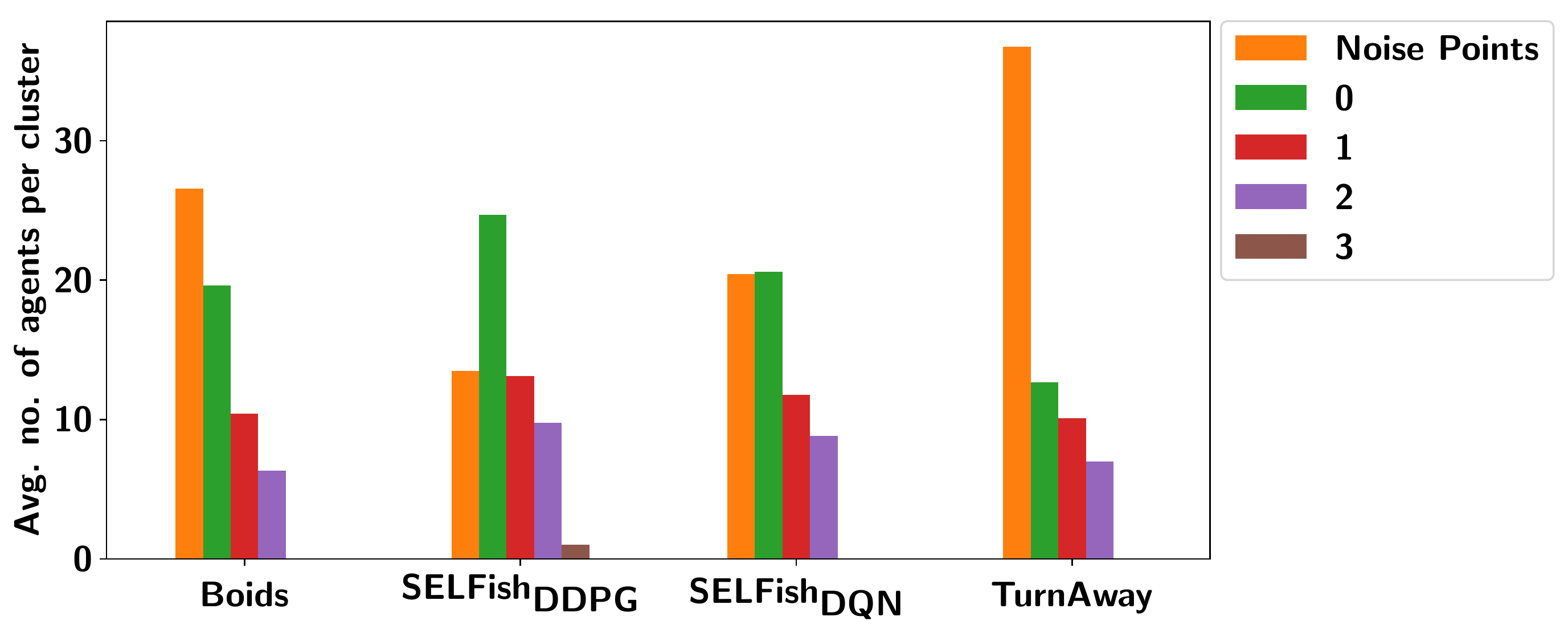}
\caption{Average number of agents in a respective cluster (cluster ID given) with noise points being agents that could not be assigned to a specific cluster.}
\label{fig:agentsPerCluster}
\end{center}
\end{figure}

Figure \ref{fig:agentsPerCluster} shows the number of agents in a specific cluster, when $ 40  $ agents were present in a space of $ 40 \times 40$ pixels.
It is visible that the TurnAway strategy produces many noise points on average.
The clusters that are found for TurnAway are mostly due to the agents moving in the same direction to avoid the predator and also overlapping when wrapping around the edges of the space. Boids and the two reinforcement learning approaches used in SELFish, DQN and DDPG, produce rather similar cluster numbers and sizes on average, with DDPG having a tendency to form one large cluster.

\begin{figure}[t]
\begin{center}
\includegraphics[width=.46\textwidth]{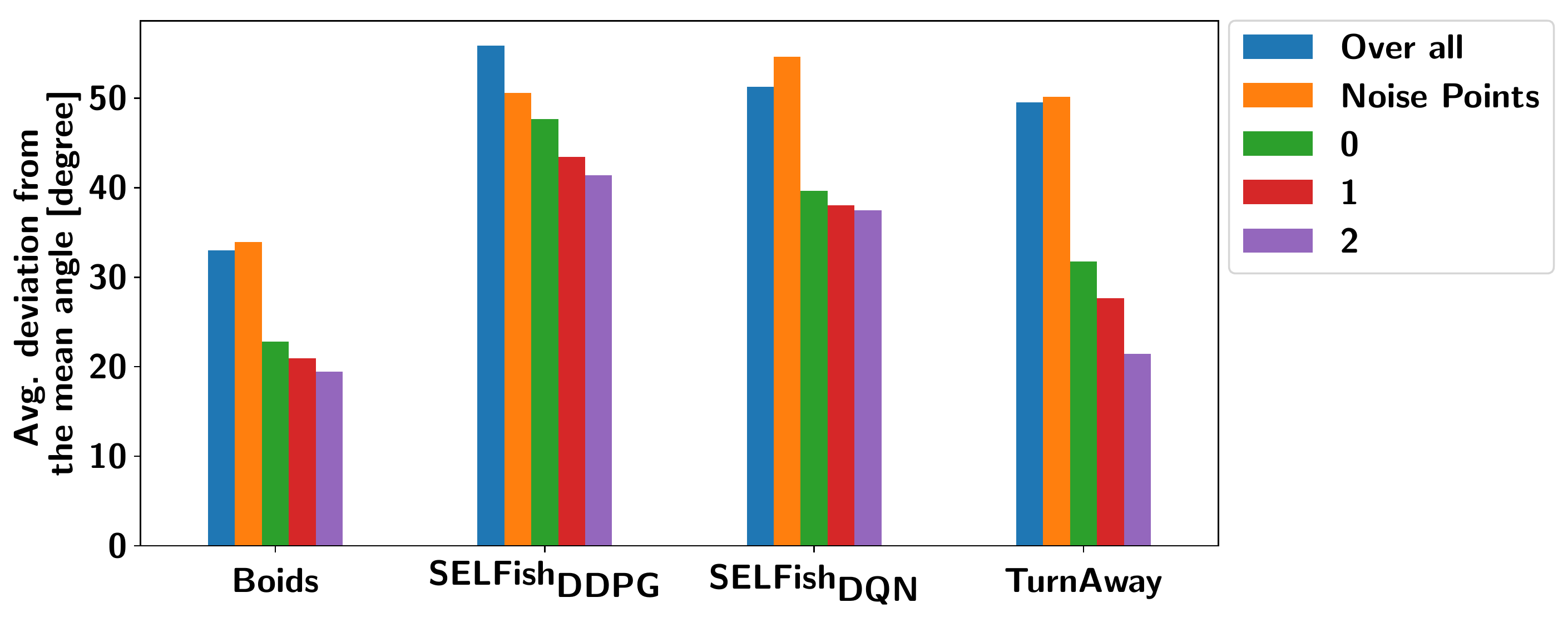}
\caption{Average deviation from the mean orientation angle of the agents over clusters.}
\label{fig:avgAngleDeviation}
\end{center}
\end{figure}

\begin{figure}[h]
\begin{center}
\includegraphics[width=.46\textwidth]{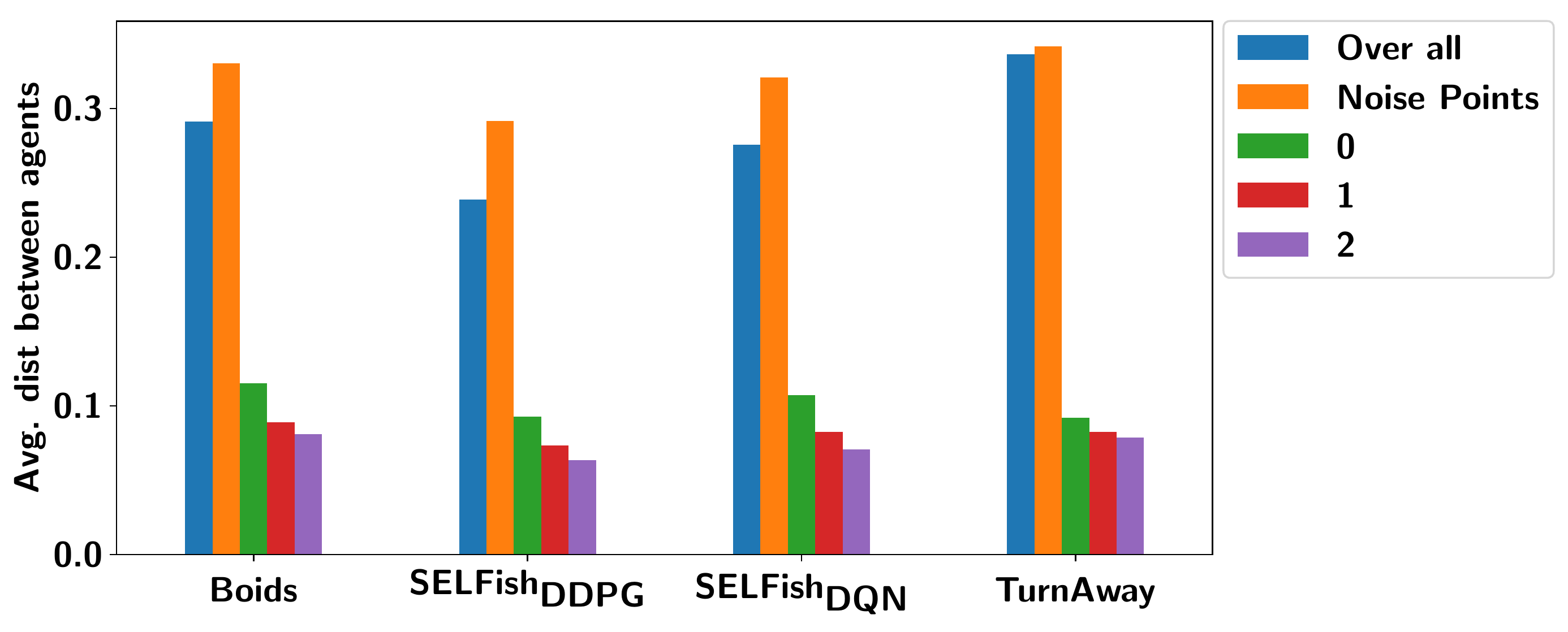}
\caption{Average pairwise distance between agents either inside clusters, between noise points or over all. Edge lengths of the space normalized to $ 1 $ for distance calculation.}
\label{fig:avgDistBetweenFishes}
\end{center}
%\vspace*{-10px}
\end{figure}

By looking at the average deviation from the mean orientation
%\vspace*{-10px}
angle of the agents inside clusters (see Figure \ref{fig:avgAngleDeviation}) one can see that Boids produces the most aligned groups of agents which generally move in the same direction.
$ \text{SELFish}_\text{DQN} $ and $ \text{SELFish}_\text{DDPG} $ are deviating more, presumably because agents following these policies tend to kind of quiver.
Also these agents show the behavior of creating a line at the point at which they would again move towards the predator because of the torus environment. % (comparable to the red lines in Figure \ref{fig:torus}).
At these lines the agents circulate until the predator moves into their direction.
For TurnAway only groups of agents moving in the same direction are detected anyway, with the average angle deviation being distorted by agents coming from the other side of the space and moving in the opposite direction.
One might question whether the swarms (respectively clusters) found for $ \text{SELFish}_\text{DQN} $ or $ \text{SELFish}_\text{DDPG} $ also solely result from the fact that the agents learned to turn away from the predator and thereby move in the same direction.
This can be countered by the observation that if the predator is pinned down at a fixed position (it cannot be removed completely as it is part of the agents' observation), the learning agents still form a swarm at the greatest possible distance from the predator where they circulate around each other.
Figure \ref{fig:avgDistBetweenFishes} shows the average pairwise distance between agents either inside clusters, between noise points or between all agents, which is homogeneous over all four agent policies, with only $ \text{SELFish}_\text{DDPG} $ tending to produce somewhat denser agent groups.
The homogeneity between the behavioral strategies with regard to the average pairwise distance also results from the DBSCAN clustering.

\subsection{Agent Survival}

\begin{figure}[]
\begin{center}
\includegraphics[width=.36\textwidth]{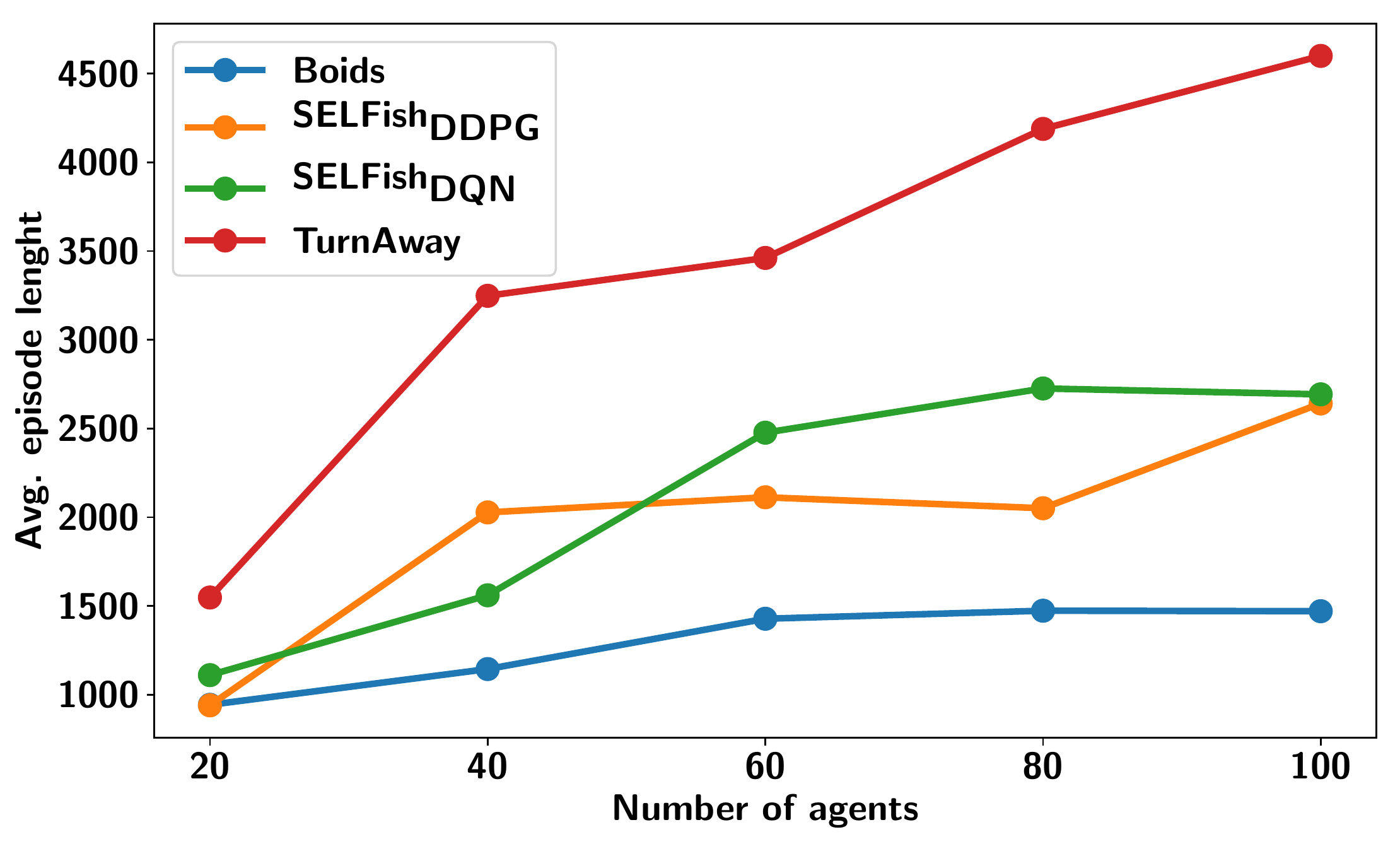}
\caption{Average episode length for each of the behavior strategies with varying number of agents in the environment.}
\label{fig:episodeLength}
\end{center}
\end{figure}

For the reinforcement learning algorithms the reward was defined such that the single learning agent received $ +1 $ for every step and $ -1000 $ for being caught. 
The maximization of the accumulated reward should encourage it to stay alive as long as possible. 
After the end of an episode, which ended when the learning agent was caught or $ 10,000 $ steps passed, the learned policy was copied to all other agents.
Figure~\ref{fig:episodeLength} shows the mean episode length for the different policies, which essentially corresponds to the mean accumulated reward of the learning agents.
For the static policies, Boids and TurnAway, it corresponds to the time it took until a certain agent was caught.
Note that although the number of agents in the environment is varied, the parameter for Boids or the policies for $ \text{SELFish}_\text{DQN/DDPG} $ are still those that were determined in smaller settings with only $ 10 $ agents.

\begin{figure}[]
\begin{center}
\includegraphics[width=.36\textwidth]{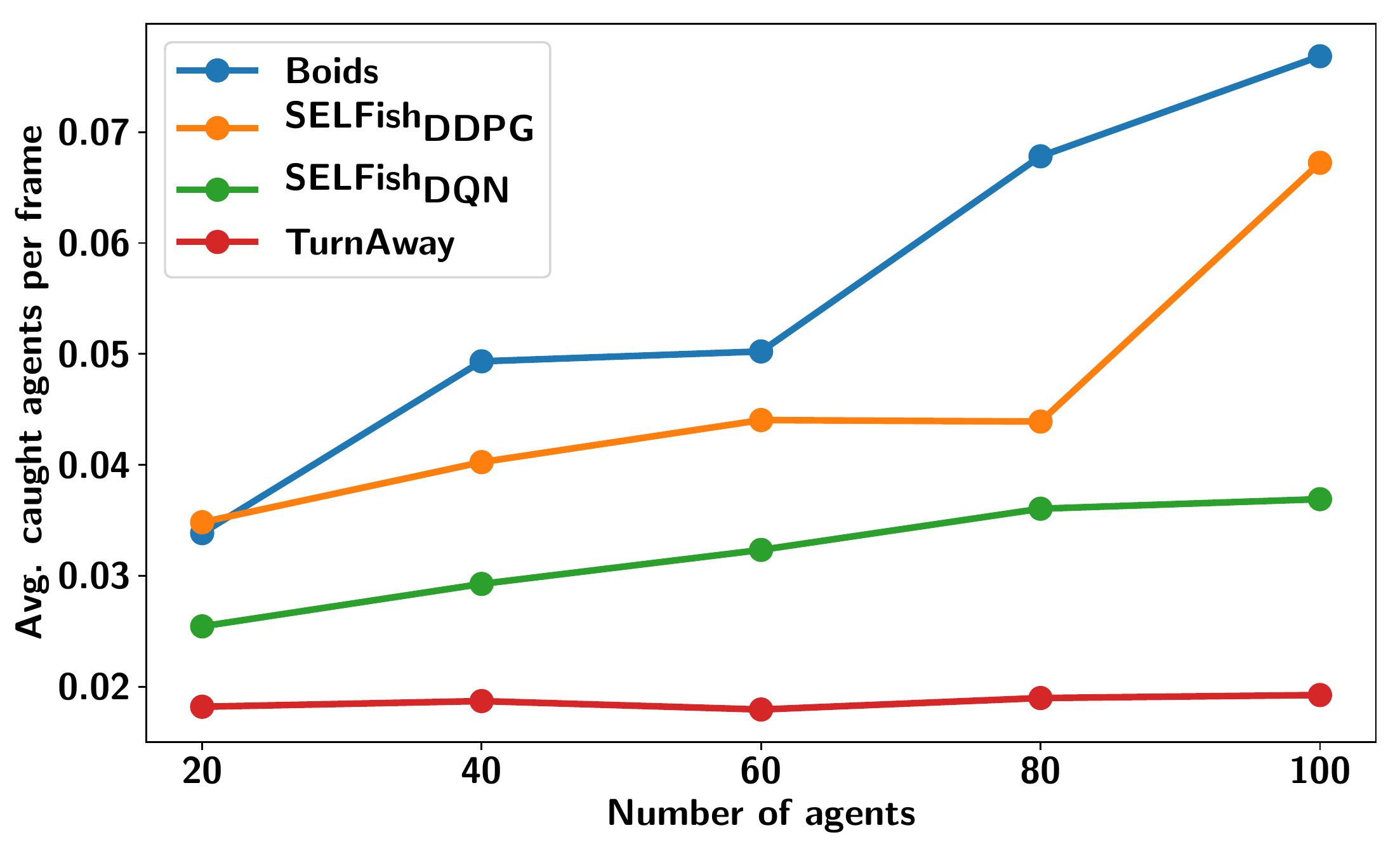}
\caption{Number of caught agents divided by the time it took with varying number of agents in the environment.}
\label{fig:caughtFishPerFrame}
\end{center}
\vspace*{-10px}
\end{figure}

It turns out that when evaluating the actual survival rate of every single agent, the best strategy to survive is to simply turn away from the predator.
This is also true considering the whole swarm, i.e. all agents.
In Figure~\ref{fig:caughtFishPerFrame}, the absolute number of caught agents in an episode was divided by the length of the episode (reduced by a transient phase of 100 frames for swarm formation).
These measurements were then again averaged over multiple episodes and runs (with different seeds).

This raises the question why this behavior was not found by the reinforcement learning algorithms.
\begin{figure}[]
\begin{center}
\includegraphics[width=.36\textwidth]{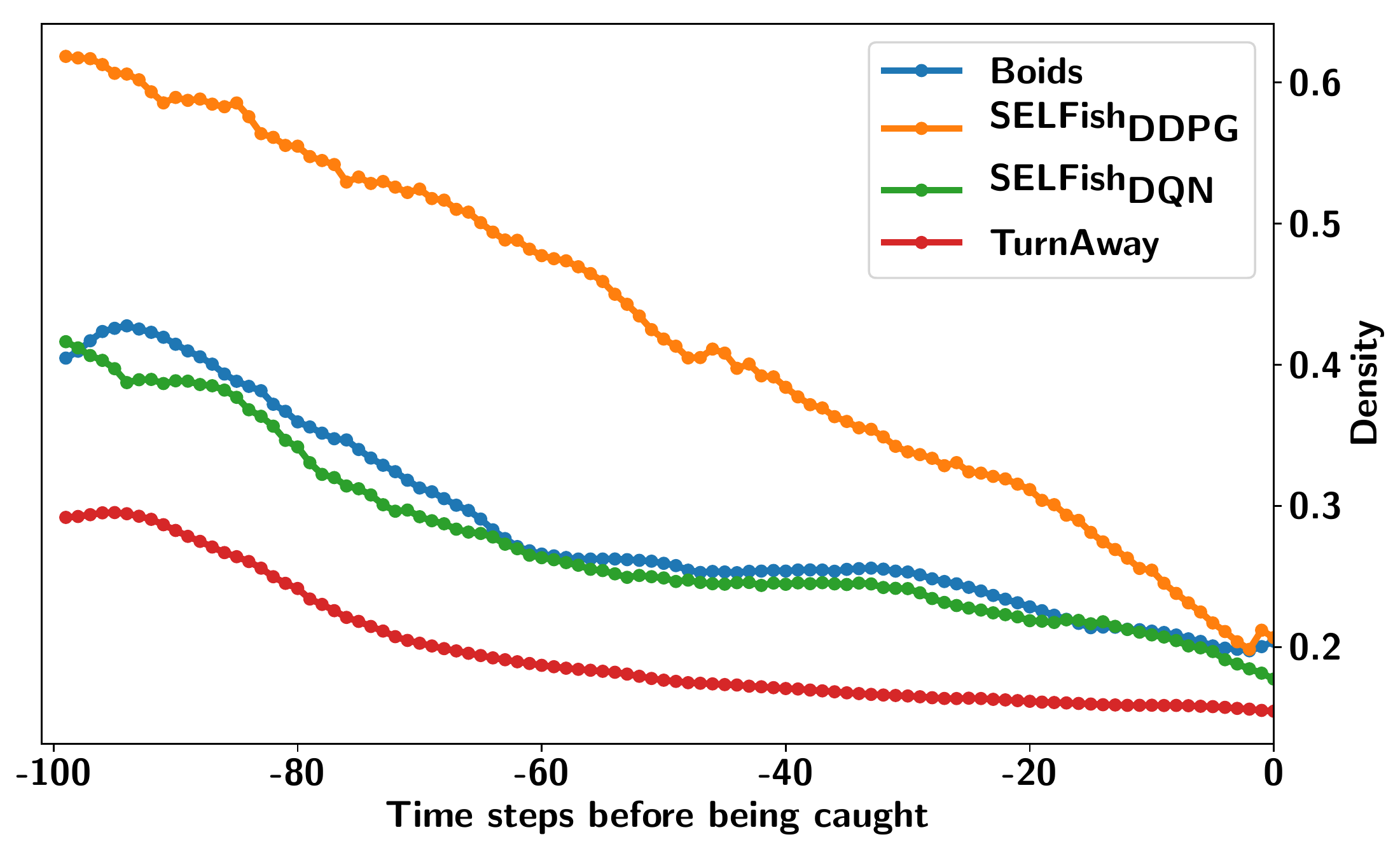}
\caption{Density of an agent in accordance to the Kernel Density Estimation in the last $ 100 $ time steps before it is caught (mean for multiple agents).}
\label{fig:densityBeforeDeath}
\end{center}
\end{figure}
The outcome of the reinforcement learning could potentially be explained considering the Prisoner's Dilemma (\cite{prisoner}).
In this game-theoretical example, prisoners $ A $ and $ B $ are kept in arrest without means to communicate.
Simultaneously, both are given the opportunity either to betray the other by testifying that the other committed the crime, or to cooperate with the other by remaining silent with the respective payoffs shown in Table~\ref{tab:prisoner}.
\begin{table}[b]
\centering
\begin{tabular}{|c|c|c|}\hline
\diagbox[width=2.5cm]{A}{ B}& \begin{tabular}[c]{@{}c@{}} $ B $ stays silent\\ (cooperates)\end{tabular}   & \begin{tabular}[c]{@{}c@{}} $ B $ betrays\\ (defects)\end{tabular} \\ \hline
\begin{tabular}[c]{@{}c@{}} $ A $ stays silent\\ (cooperates)\end{tabular} &\diagbox[width=2.3cm]{-1}{-1} &\diagbox[width=2.3cm]{-3}{0}  \\ \hline
\begin{tabular}[c]{@{}c@{}} $ A $ betrays\\ (defects)\end{tabular}& \diagbox[width=2.3cm]{0}{-3} & \diagbox[width=2.3cm]{-2}{-2}  \\ \hline
\end{tabular}
\caption{Prisoner's dilemma payoff matrix}
\label{tab:prisoner}
\end{table}
\begin{figure*}[!t]
\centering
\includegraphics[width=.9\textwidth]{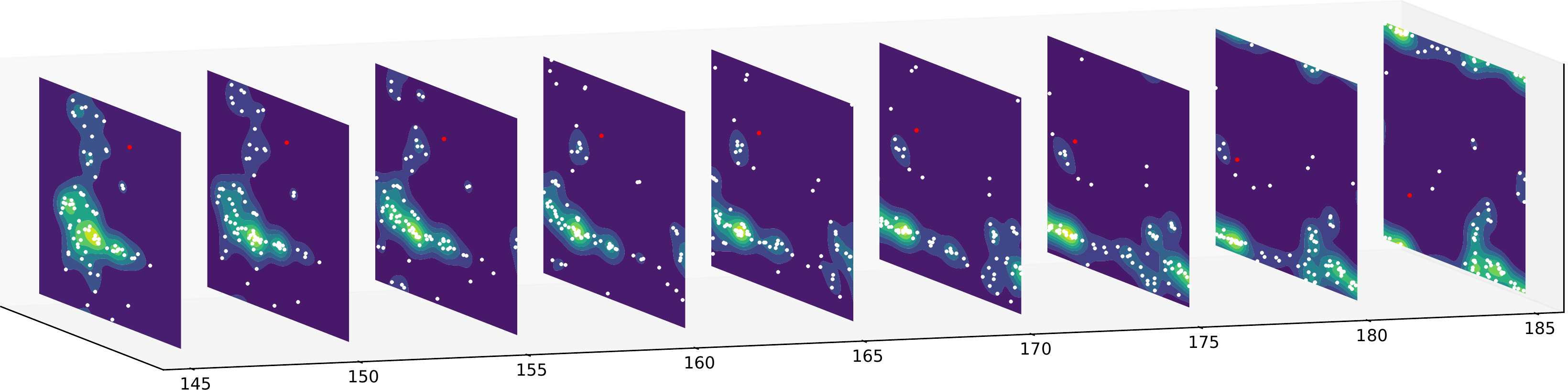}
\caption{Separation of agents from the swarm before being caught.}
\label{fig:DQN-SwarmSeparation}
%\vspace*{-5px}
\end{figure*}
The only Nash equilibrium (\cite{nash}) is that both prisoners defect as this yields less charge for each of them than if one stays silent while the other prisoner keeps its strategy unchanged and testifies that the other committed the crime (betrays).
The dilemma is that mutual cooperation yields a better outcome although it is not rational from a self-interested perspective.
For our reinforcement learning setting it could be the case that the TurnAway strategy was not found because the learning process got stuck in the Nash equilibrium of staying with the swarm (analogous to the mutual defection in the Prisoner's Dilemma).
If all agents keep their policy of staying close to each other, the one agent deviating has a higher chance of being chosen as prey.
Our learning procedure is in conformity with this as one learning agent adjusts its policy in such a way that it obtains the highest reward while the policies of the other agents stay unchanged (during an episode).
This assumption is also supported by looking at the procedure how agents are caught (see Figure \ref{fig:DQN-SwarmSeparation}):
When the predator moves in the direction of the swarm, it collaboratively moves away, with a few agents being left behind. 
The community of the agents gets smaller and smaller as some sheer off until one is separated and picked as prey.
This is also evident considering the density measurements of agents in the time steps before it is being caught.
Figure \ref{fig:densityBeforeDeath} shows the density around an agent in accordance to the Kernel Density Estimation (cf. Figure \ref{fig:DQNSwarmFormation} and \ref{fig:DQN-SwarmSeparation}) in the last $ 100 $ time steps of its life.

\section{Conclusion and Future Work}
With SELFish we showed that flocking behavior can emerge solely from the fact that agents trained by multi-agent reinforcement learning try to avoid being caught by a predator, given the circumstance that flocking yields a benefit like distracting the predator.
Only one agent was trained at a time with a reward structure that encourages to avoid being caught as long as possible. 
After each episode the learning policy was copied onto all other agents.
The results for $ \text{SELFish}_\text{DQN} $ and $ \text{SELFish}_\text{DDPG} $ concerning the alignment and cohesion but also with regard to the survival chances of the agents were compared with Boids, a common approach for algorithmic flocking simulations.
Out results show, that the measurements for the swarm are quite similar to Boids.
Considering the survival of an agent, surprisingly, the reinforcement learning algorithms did not find the policy of simply turning away from the predator (without caring about flocking) although it yields higher accumulated rewards w.r.t our reward structure.
We propose that staying in the swarm is a Nash equilibrium (comparable to defecting in the Prisoner's dilemma) and want to further investigate this assumption.
Also, we would like to examine if other beneficial properties of a swarm, like increased hydrodynamic efficiency or easier search for food, which were not modeled by us, also lead to flocking behavior in a reinforcement scenario.
This would probably also facilitate the steering of the swarm.
Co-evolution of the behavior of the predator and its prey through reinforcement learning could be further investigated in our continuous environment.
In our setting, agents could freely roam in a torus-like environment without obstacles or collisions.
Naturally, there are enhancements to this like adding walls, obstacles and collisions between the agents.
%setting, kollisionen, pyhisk
%andere anreize, wie futter, dynamik,
%coevolution des predators in kontiniuerlicher domäne
%untersuchung nash equip turn away

%\begin{figure}[t]
%\begin{center}
%\includegraphics[width=2.1in,angle=-90]{fig/fig1.eps}
%\caption{``Energies'' (inferiorities) of strings in a first-order
%  phase transition with latent heat $\Delta\epsilon$.}
%\label{fig1}
%\end{center}
%\end{figure}

~\\

\footnotesize
\bibliographystyle{apalike}
\bibliography{bibliography} % replace by the name of your .bib file

\end{document}